# Phase-field study on the segregation mechanism of Cr to lamellar interface in C40-NbSi$_2$/C11$_b$-MoSi$_2$ duplex silicide

Toshihiro Yamazaki[1], Yuichiro Koizumi[2], Akihiko Chiba[2], Koji Hagihara[3], Takayoshi Nakano[4], Koretaka Yuge[5], Kyosuke Kishida[5] and Haruyuki Inui[5]

[1]Department of Materials Processing, Graduate School of Engineering, Tohoku University, 6-6-02 Aoba Aramaki, Aoba-ku, Sendai, Miyagi 980-8579, Japan

[2]Institute for Materials Research, Tohoku University, 2-1-1 Katahira, Aoba-ku, Sendai, Miyagi 980-0011, Japan

[3]Department of Adaptive Machine Systems, Graduate School of Engineering, Osaka University, 2-1 Yamada-oka, Suita, Osaka 565-0871, Japan

[4]Division of Materials and Manufacturing Science, Graduate School of Engineering, Osaka University, 2-1 Yamada-oka, Suita, Osaka 565-0871, Japan

[5]Department of Materials Science and Engineering, Kyoto University, Sakyo, Kyoto 606-8501, Japan

**Abstract**

Cr-segregation to a lamellar interface in NbSi$_2$/MoSi$_2$ duplex silicide has been examined by a newly developed phase-field model. The model can take into account the segregation energy evaluated by a first principles calculation to reflect the chemical interaction between solute atoms and the interface in addition to the elastic interaction. Cr segregation occurs at the interface in the case with segregation energy whereas no segregation occurs in the case with only elastic interaction.  However, the segregation is much smaller than that observed in the experiment when the segregation energy was evaluated by the first principles calculation without lattice vibration (i.e. for 0 K). Another simulations with the segregation energy with lattice vibration results in segregation comparable to that in the experiment.   Thus, it has been revealed that the solute-interface chemical interaction and its temperature dependence is responsible for the interfacial segregation of Cr.

1. Introduction

Due to its high melting temperature, low density, good thermal-conductivity [1], $MoSi_2$ with $C11_b$-structure (Fig.1a) has attracted attention as refractory material for gas power generation systems at ultrahigh-temperature where Ni-based superalloys are difficult to overcome.  Monolithic $MoSi_2$ exhibits good plastic deformability at low-temperature owing to the operation of several slip systems, but its fracture toughness at low-temperature and strength at high-temperature is insufficient [2-4]. There have been many studies aiming at the improvement of the mechanical properties through forming composites with other material.  For instance, the composites with ceramics such as $Al_2O_3$ and SiC exhibit improved high temperature strengths, but lack the thermal shock resistance and low-temperature toughness [5,6].  Reinforced by combining with another silicide is attractive.  Disilicides with C40 structure (Fig.1b) are attractive for reinforcing $MoSi_2$ matrix since they exhibit anomalous strengthening at higher temperatures than $MoSi_2$ does [7-10].  Duplex silicides composed of C40-$NbSi_2$ and $C11_b$-$MoSi_2$ are   one of the most promising candidates for practical use which requires further improvement of mechanical properties [11].

The $NbSi_2$/$MoSi_2$ duplex silicide shows anomalous strength at temperatures between 800˚C and 1500˚C [12].  In a previous study, a well-oriented lamellar structure composed of C40-type Nb-rich phase and $C11_b$-type Mo-rich phase has been formed in the alloy with the nominal composition of $Mo_{0.85}Nb_{0.15}Si_2$ by unidirectional solidification followed by annealing at dual phase temperatures.  Actually, it was found that the lamellar structure improves the mechanical properties of duplex-silicide such as low-temperature toughness and high-temperature strength [13-16].   The lamellar structure is relatively thermally stable, and it can remain even after annealing at 1673 K for 168 h for instance [15]. However, lamellar structure can coarsen and collapse after a prolonged annealing of 336 h [17]. Therefore, the thermal stability of lamellar structure needs to be further improved for practical use. The addition of 1at% Cr to $NbSi_2/MoSi_2$ duplex silicide was found to improve the thermal stability of C40/C11$_b$ lamellar structure. The lamellar structure of $(Mo_{0.85}Nb_{0.15})_{0.97}Cr_{0.03}Si_2$ alloy does not collapse even after the prolonged annealing [17]. The improved thermal stability of lamellar structure has been believed to be due to the decrease in the lattice misfit at the C40/C11$_b$ lamellar interface by Cr-segregation. The Cr-segregation at the interface has already been proved by scanning transmission electron microscopy and energy dispersive spectroscopy. This Cr-segregation is suggested to change the lattice parameter in the vicinity of lamellar interface to reduce the lattice misfit and improve the planarity and the thermal stability of lamellar structures. However, the suggested mechanism of Cr-segregation at the C40/C11$_b$ interface has not been proved yet. Also, the quantitative relationship between Cr-segregation and lattice-misfit change is unclear. Furthermore, it is also to be clarified whether Cr is the optimum additive element to stabilize the lamellar structure in the $NbSi_2/MoSi_2$ duplex silicide.

Phase-field method is a versatile technique for simulating microstructural evolution and interfacial segregation in various materials on the basis of thermodynamics and diffuse interface concept. So far, we have been applying the phase-field method to the studies of microstructural evolution and segregation at interfaces such as stacking faults [18], twin boundary [19] and anti-phase boundaries [20]. Especially, this method has been used for quantitative investigations of interfacial segregation which are difficult to be conducted experimentally. The objective of this study is to clarify factors dominating the evolution of lamellar structure and the interfacial segregation mechanism of additive elements in $NbSi_2$/$MoSi_2$ duplex silicide by phase-field method.

## 2. Method

### 2.1. Phase-field model

As illustrated in Fig. 1a and b, both C40 hexagonal structure and $C11_b$ tetragonal structure consists of two sublattices, i.e. transition metal (TM) lattice (represented by red sphere) and Si-sublattice (blue spheres) [21]. It is assumed that no antisite atoms are formed, i.e. Si atoms cannot occupy TM-site and vice versa for simplicity. Although small amount of Si-atoms can occupy TM-site, the amount is negligibly small. C40- and $C11_b$- structures may appear quite different from each other, but the atomic arrangement within an atomic layer on (0001) plane of C40 phase and that on (110) plane of $C11_b$-phase are nearly identical (Fig. 1c, d). The difference between the two crystal structures is the sequence of stacking layers. Namely, C40-structure has ABDABD-type stacking whereas $C11_b$-structure has ACAC-type stacking. Coherent lamellar interfaces are formed parallel to $(0001)_{C40}$ and $(110)_{C11b}$ planes [22, 23]. Six types of lamellar interfaces with different interconnecting layers can be formed. The previous first principles calculation study elucidated that the interface with the stacking sequence of …DABDAB|ACAC…. has the lowest interfacial energy.

Fig. 2a illustrates the atomic arrangement of the lamellar interface with the lowest energy. The order parameter $\phi$ is defined so that $\phi = 0$ represent C40-phase and $\phi = 1$ represents $C11_b$-phase (Fig. 2b). In order to distinguish the ACA-stacking of C40-structure and the ABD-stacking $C11_b$-structure, the stacking of at least three layers needs to be taken into account. For the D-layer indicated by the green arrow, the third layer (counted from itself) in the interface-side is B-layer which in within ABD-stacking of C40-phase. But for the A-layer indicated by the cyan arrow, the third layer (i.e. the A-layer indicated by the purple arrow) is no more within the C40-structure but in the $C11_b$-structure. Accordingly, we consider the A-layer with the cyan arrow is in an intermediate state between C40-structure and $C11_b$-structure. Similarly, the B-layer with the blue arrow is also considered to be in an intermediate state. In the same manner, the A-layer with the purple arrow and the C-layer with the magenta arrow in the $C11_b$-side are considered to be in intermediate state. The layers in the intermediate states are considered to be within the diffuse interface region. The D-layer with the green arrow and the A-layer with the red arrow are considered to be at the edges of the diffuse interface region. Accordingly, the $C40/C11_b$ diffuse interface with the thickness of 6 atomic layers is defined as a region where the order parameter changes from 0 to 1 continuously.

*2.2. Alloy composition and initial condition*

The alloy composition of duplex silicide selected is $(Mo_{0.862}Nb_{0.138})_{0.97}Cr_{0.03}Si_2$. This composition is determined as follows. First, equilibrium compositions of C40- and $C11_b$-phases were calculated for a pseudo-binary $(Mo_{0.85}Nb_{0.15})Si_2$ alloy. Equal volumes of $C11_b$-phase and C40-phase with the equilibrium concentration are connected. Then, 1 at% of Cr atoms is substituted uniformly to the TM-sites in both phases keeping the ratio of Mo-concentration and Nb-concentration in each phase. This resulted in the composition of $(Mo_{0.862}Nb_{0.138})_{0.97}Cr_{0.03}Si_2$.

*2.3. Free energy*

In the phase-field simulation, the total free energy of the system needs to be described as a functional of order parameter and solute concentrations which are functions of position. We take into account the chemical free energy, the gradient energy and the elastic strain energy as the components of the total free energy. Details of each component of the energies are described below.

*2.3.1. The chemical free energy*

The chemical free energy densities of C40-phase and $C11_b$-phase are evaluated by the CALPHAD method [24] with the thermodynamic data in the literature [21, 25].

Since there is no thermodynamic data of Mo-Si-Cr system, they are estimated from $MoSi_2$-$CrSi_2$ pseudo binary phase diagram [26] by the Thermo-Calc software. First, the energy densities of C40 phase ($f_{C40}$) and $C11_b$ phase ($f_{C11b}$) are calculated as functions of temperature and composition using the CALPHAD method. Then, the chemical energy density $f_{chem}$ of the intermediate state is defined as

$$f_{chem}(c_M, \phi) = f_{C40}(c_M)h(\phi) - f_{C11b}(c_M)(1-h(\phi)) + W_{interface} \cdot g(\phi) \qquad (1)$$

where $h(\phi)$ is the interpolation function and $g(\phi)$ is the double-well function. The interpolation function $h(\phi)$ needs to increase monotonically with increasing $\phi$ from 0 to 1 satisfying the conditions of $h(0)=0$, $h(1)=1$ and $(dh/d\phi)_{\phi=0} = (dh/d\phi)_{\phi=1} = 0$. The double well function $g(\phi)$ needs to have two minima of $g(0)=0$ and $g(1)=0$. The first two terms of Eq. (1) is for interpolating the energy of C40-phase and $C11_b$-phase. The third term of Eq.(1) is for penalizing the intermediate state. The $W_{interface}$ is a parameter which determines the magnitude of the energy penalty. There are several types of functions for $h(\phi)$ and $g(\phi)$ suggested. In this study, the following functions are employed.

$$h(\phi) = \phi - \frac{\sin(2\pi\phi)}{2\pi} \qquad (2)$$

$$g(\phi) = \frac{1-\cos(2\pi\phi)}{2\pi} \qquad (3)$$

The value of $W_{interface}$ can be derived from the interfacial energy $\gamma$ and the thickness of diffuse interface $d$ as $W_{interface} = \gamma/d$. In the present study, the value of $W_{interface}$ is determined to be 171 J/mol from $\gamma = 27$ mJ/m$^2$ which is derived from the first principles calculation and $d=1.3$ nm which is the thickness of 6 atomic layers. The chemical free energy is plotted in Fig. 3 as a function of $\phi$ and Nb-concentration with the concentrations of Si and Cr fixed at $c_{Si} = 66.7$ at% and $c_{Cr} = 1.0$ at%, respectively. The free energy curves of C40-phase ($\phi=0$) and $C11_b$-phase ($\phi=1$) are smoothly interpolated by the function $h(\phi)$ with a hump corresponding to the term $W_{interface} \cdot g(\phi)$ in Eq.(1).

*2.3.2. The gradient energy*

The gradient energy is the excess free energy associated with the inhomogeneities of the order parameter and solute compositions. Here Mo is considered to be solvent while Nb and Cr are solute, and Nb- and Cr-concentrations ($c_{Nb}$ and $c_{Cr}$) are defined as independent variables) and Mo-concentration ($c_{Mo}$) is dependent variable. (Si concentration is fixed to be 2/3). Then, the gradient energy density is given by

$$f_{grad}(\nabla c_{Nb}, \nabla c_{Cr}, \nabla \phi) = \kappa_{Nb}(\nabla c_{Nb})^2 + \kappa_{Cr}(\nabla c_{Cr})^2 + \kappa_{NbCr}\nabla c_{Nb}\nabla c_{Cr} + \kappa_{\phi}(\nabla \phi)^2 \qquad (4)$$

where $\kappa_{\phi}$ and $\kappa_{M}$, $\kappa_{MM'}$ ($M \neq M'$, $M$ = Nb, Cr) are the gradient energy coefficients for the order parameter, concentration of solute $M$ and the product of concentration gradients of different solute elements, respectively. The value of $\kappa_{\phi}$ is derived to be $2.66 \times 10^{-11}$ J/m from the interfacial energy (27.2 mJ/m$^2$) and the interfacial thickness (1.3 nm) on the basis of equilibrium-interface condition. The evaluation of $\kappa_{M}$ and $\kappa_{MM'}$ are generally problematic. The values of the gradient energy coefficients could be derived a priori from the definition if the free energy density of non-uniform system was described as a function of the gradients. However, the description of the free energy used in the CALPHAD method, which is used in the present study, is applicable to only uniform systems. Therefore, in most of phase-field models, gradient energy coefficients for solute concentration is neglected assuming that the interfacial energy can be attributed to the special change in order parameter across interface rather than that of composition change. Following the previous studies, the values of $\kappa_{M}$ and $\kappa_{MM'}$ are assumed to be zero, i.e. gradient energies for solute concentrations are neglected in this study.

*2.3.3 The elastic strain energy*

The elastic strain energy arising from the lattice mismatch between C40- and C11$_b$ -structures is described based on Hooke's law as in the following equations.

$$f_{\text{str}}(c_{\text{Nb}}, c_{\text{Si}}, c_{\text{Cr}}, \phi) = \frac{1}{2} \int C_{ijkl} \varepsilon_{ij}^{\text{el}} \varepsilon_{kl}^{\text{el}} dV \qquad (5)$$

$$\varepsilon_{ij}^{\text{el}} = \varepsilon_{ij}^{\text{total}} - \varepsilon_{ij}^{\text{eigen}}(c_{\text{Nb}}, c_{\text{Cr}}, \phi) \qquad (6)$$

$$\varepsilon_{ij}^{\text{eigen}}(c_{\text{Nb}}, c_{\text{Cr}}, \phi) = \eta_{ij,\text{Mo}} c_{\text{Mo}} \phi + \eta_{ij,\text{Nb}} c_{\text{Nb}} \phi + \eta_{ij,\text{Cr}} c_{\text{Cr}} \phi. \qquad (7)$$

$C_{ijkl}$ is the elastic constants, $\varepsilon_{ij}^{\text{el}}$ the elastic strain, $\varepsilon_{ij}^{\text{total}}$ the total strain, $\varepsilon_{ij}^{\text{eigen}}$ the eigen strain and $\eta_{ij}$ the lattice mismatch. Homogeneous elasticity is assumed (i.e. elastic modulus difference between C40- and C11$_b$-phases is neglected) for simplicity, and the elastic constants of C11$_b$-MoSi$_2$ [27] are employed for the values of $C_{ijkl}$. The $\varepsilon_{ij}^{\text{eigen}}$ is defined as a linear function of the order parameters $\phi$ and solute concentrations assuming Vegard's law. The lattice mismatch for transition metal $M$ is defined using lattice sizes of disilicides $M$Si$_2$ as below

$$\eta_{ij,M} = \begin{cases} (l_{i,M\text{Si}_2}^{\text{C11}_b} - l_{i,M\text{Si}_2}^{\text{C40}})/l_{i,M\text{Si}_2}^{\text{C40}} & \text{for } i = j \\ 0 & \text{for } i \neq j \end{cases} \qquad (8)$$

Lattice sizes $l_1$, $l_2$ and $l_3$ of disilicides (MoSi$_2$, NbSi$_2$, CrSi$_2$) are derived from the lattice parameters of C40- and C11$_b$-structures evaluated by the first-principles calculation. Lattice parameters include those of fictitious disilicides which are difficult to have C40- or C11$_b$-structure thermodynamically, and do not exist in reality. The ability to evaluate such values is the advantage of the use of first principles calculation. Lattice sizes of C40- and C11$_b$-disilicides of each transition metals are evaluated by defining equivalent supercells in the two crystal structures as.

$$\begin{aligned} l_1^{\text{C40}} &= 2\sqrt{3}a_{\text{C40}}, & l_1^{\text{C11}_b} &= 2c_{\text{C11}_b} \\ l_2^{\text{C40}} &= a_{\text{C40}}, & l_2^{\text{C11}_b} &= \sqrt{2}a_{\text{C11}_b} \\ l_3^{\text{C40}} &= 2c_{\text{C40}}, & l_3^{\text{C11}_b} &= 3\sqrt{2}a_{\text{C11}_b} \end{aligned} \qquad (9)$$

Lattice parameters and lattice sizes of the transition metal disilicides are liseted in Table 1. Lattice sizes in all the three directions of CrSi$_2$ are smaller than those of MoSi$_2$ and NbSi$_2$ for both C40- and C11$_b$-structures.

*2.4. Time evolution*

The total energy of the system, $G_{sys}$, is given by integrating the summation of the chemical free energy, the gradient energy and the elastic strain energy as

$$G_{sys} = \int \left[ f_{chem}(c_{Nb}, c_{Si}, c_{Cr}, \phi) + f_{grad}(c_{Nb}, c_{Si}, c_{Cr}, \phi) + f_{str}(c_{Nb}, c_{Si}, c_{Cr}, \phi) \right] dV. \quad (10)$$

$G_{sys}$ is the functional of solute concentrations and the order parameter which are the function of position. Solute concentrations of $c_{Nb}$ and $c_{Cr}$ are independent variable since $c_{Si}$ is fixed to be 2/3 as mentioned above. Mo-concentration can be treated as a dependent variable of $c_{Nb}$ and $c_{Cr}$ due to the relation of $c_{Mo} = 1/3 - c_{Nb} - c_{Cr}$. The spatial distributions of $c_{Nb}$, $c_{Cr}$ and $\phi$ are evolved so as to decrease the total energy of the system ($G_{sys}$) by solving the Cahn-Hilliard equation [28] for $c_{Nb}$ and $c_{Cr}$, and solving the Allen-Cahn equation [29] for $\phi$ simultaneously. The Cahn-Hilliard equation for multi-element system is describes as

$$\frac{\partial c_M}{\partial t} = \left[ \frac{D_M c_M}{kT} \left\{ (1-c_M) \nabla \left( \frac{\delta G_{sys}}{\delta c_M} \right) - \sum_{M' \neq M} c_{M'} \nabla \left( \frac{\delta G_{sys}}{\delta c_{M'}} \right) \right\} \right] \quad (M, M' = \text{Nb}, \text{Cr}) \quad (11)$$

where $D_M$ is the self-diffusion coefficient of element $M$. The diffusion coefficients of transition metal atoms are assumed to be equal to that of Mo self-diffusion along a-axis in C11$_b$-MoSi$_2$ which is available in literature [30]. The Allen-Cahn equation are respectively expressed as,

$$\frac{\partial \phi}{\partial t} = -\alpha \left( \frac{\delta G_{sys}}{\delta \phi} \right) \quad (12)$$

where $\alpha$ is the mobility for transformation between C40-phase and C11$_b$-phase.

2.5. Simulation condition

2.5.1 Simulation of interface relaxation without segregation energy

As initial condition, equal volumes of C40-phase with the composition of (Mo$_{0.968}$Nb$_{0.032}$)$_{0.97}$Cr$_{0.03}$Si$_2$ and C11$_b$-phases with that of (Mo$_{0.756}$Nb$_{0.244}$)$_{0.97}$Cr$_{0.03}$Si$_2$ are connected with a sharp interface at the center of the system. The total alloy composition is (Mo$_{0.862}$Nb$_{0.138}$)$_{0.97}$Cr$_{0.03}$Si$_2$ as mentioned above. The compositions of each phase are determined as follows. First, equilibrium compositions of C40-phase and C11$_b$-phase for the pseudo-binary alloy with the total composition of (Mo$_{0.862}$Nb$_{0.138}$)Si$_2$ are calculated to be (Mo$_{0.756}$Nb$_{0.244}$)Si$_2$ and (Mo$_{0.968}$Nb$_{0.032}$)Si$_2$, respectively. Then, 1 at% of transition metal atoms (i.e. Mo and Nb) are replaced with Cr, keeping the ratio of $c_{Nb}/c_{Mo}$. The system simulated is 24 nm long, and a one-dimensional simulation box with 256 grid points is used. The periodic boundary condition is employed at the edges of the grid for calculating the elastic strain energy. Because of the periodic boundary condition another interface is formed on the edges of the grid. The profiles of the order parameter and solute concentrations are symmetric about the center of the two interfaces. Therefore, only the middle half of the system will be displayed in the results. Fig. 4 shows the profiles of the order parameter and solute concentrations for the initial condition. The distributions of the order parameter and solute concentrations are relaxed until an equilibrium is reached at 1673 K. Simulations are conducted in the two cases where the elastic strain energy is taken into account or not. The effect of the relaxation of the lattice misfit will be examined by comparing the results of the two cases.

*2.5.2. Simulation of interfacial migration*

In the experiment, the C40/C11$_b$ lamellar structure is formed by the precipitation of C11$_b$-lamellae from C40-matrix. It is possible that Cr segregation occur during the course of transformation from C40 single phase to C40 + C11$_b$ duplex phase in the cooling process. According to MoSi$_2$-NbSi$_2$ pseudo-binary phase diagram [26], C40- and C11$_b$-phases coexist at temperatures below 2150 K. Volume fraction of Mo-rich C11$_b$-phase which grows from C40-phase increases with decreasing temperature, and these phases have almost equal volume fraction in equilibrium at the simulation temperature of 1673 K. In order to examine the effect of solute redistribution during the growth of C11$_b$-phase on the segregation, simulations of interface migration have been conducted. In these simulations, initially, a small block of the $C11_b$-phase with the size of 24 grid points (2.25 nm) and equilibrium concentration is embedded in C40-matrix supersaturated with Mo. The Mo-supersaturation gives the driving force for the growth of the $C11_b$-phase (i.e. interfacial migration). The initial concentration of C40-phase is determined so that the total composition of the system is same as that in the simulation of static interfacial segregation described above. The interface migration and during the course of equilibration is examined focusing on the changes in the solute distribution. In the simulation of interface migration, the diffusion coefficient of Cr-atom is assumed to be variously lower than that of Mo-atom because of the interaction between solute atoms and the migrating interface.

*2.5.3 Simulation of interface relaxation without segregation energy*

In the simulations described above, the diffuse interface are defined as a region where the local state monotonically changes from C40-phase to $C11_b$-phase with the energy penalty given by the double well potential, as is often the case with conventional phase-field simulations. In such cases, chemical interaction between the interface and solute atoms are not taken into account, and segregation by chemical solute-boundary interaction cannot be simulated. Here an improved model is developed by taking into account the interaction between the additive solute atom and the atomic layers located within the diffuse interface region. In particular, in order to take into account the stability of additive elements at each atomic layer around the interface, segregation energies are evaluated on the basis of the first-principles calculation. The procedure for the first principles calculation of segregation energy is detailed in a separated paper [33]. Here the procedure used for determining the segregation energy is briefly explained. First, a superlattice with the interface as depicted in Fig. 2a is formed. Then, the energies of superlattices which have the additive Cr-atoms at the TM-sites on one of the 6 atomic layers within the diffuse interface are computed by the first-principles calculation. The energy of the superlattice in which the Cr-atom is located on the layer at the C40-matrix side end of the diffuse interface (i.e. the layer-g in Fig. 2a) is used as the reference state energy. In order to use the discrete data of the segregation energy in the continuum model of phase-field method, a continuous function of order parameter $\varphi$ which describes the energy change due to the change in the location of Cr-atom within the diffuse interface is needed. As such a continuous function, the segregation energy function is defined as in the followings.

First, the energy of Cr-atom in C40-phase ($\phi=0$) and that in C11$_b$-phase ($\phi=1$) are defined 0 and $\Delta e_{\phi=1.0}$, respectively. Then, a function $E_{\text{Cr}}^{\text{base}}(\phi)$ is defined by interpolating these energies simply using the interpolation function $\{1-\cos(\pi\phi)\}/2$ as

$$E_{\text{Cr}}^{\text{base}}(\phi) = \Delta e_{\phi=1.0}\{1-\cos(\pi\phi)\}/2. \tag{13}$$

The $E_{\text{Cr}}^{\text{base}}(\phi)$ is used as the baseline for describe the energy of Cr-atom within the diffuse intereface where $0<\phi<1$. The segregation energy function $E_{\text{Cr}}^{\text{Seg}}(\phi)$ is described as

$$\begin{aligned}E_{\text{Cr}}^{\text{Seg}}(\phi) &= \Delta e_{\phi=0.2}\exp\left(\frac{-(\phi-0.2)^2}{b}\right)+\Delta e_{\phi=0.4}\exp\left(\frac{-(\phi-0.4)^2}{b}\right)+\Delta e_{\phi=0.6}\exp\left(\frac{-(\phi-0.6)^2}{b}\right)\\ &+\Delta e_{\phi=0.8}\exp\left(\frac{-(\phi-0.8)^2}{b}\right)+\Delta e_{\phi=1.0}\left(\frac{1-\cos(\pi\phi)}{2}\right)\end{aligned} \tag{14}$$

where $\Delta e_{\phi=p}$'s ($p$ = 0.2, 0.4, 0.6, 0.8) are the deviation from $E_{\text{Cr}}^{\text{base}}(\phi)$ of the segregation energy of Cr-atom located at the 2nd, 3rd, 4th and 5th layer from the C40 side edge ($\phi=0$), and $b$ is the constant which determine the spread of the Gaussian functions. The Gaussian functions are used as a simple form of function which connect the segregation energies smoothly. Figure 5 depicts the segregation energy and the segregation energy function. The black points show segregation energies at the 6 atomic layers evaluated by the first principles calculations, and each segregation energy for peak value shown by orange solid line in Fig. 5a.

The segregation energy $E_{\text{Cr}}^{\text{Seg}}(\phi)$ given by Eq. (14) is for a unit concentration of solute atom. Since the local excess energy due to the segregation should be proportional to the solute local concentration, the product of the segregation energy function and Cr-concentration (i.e. $E_{Cr}^{Seg}(\phi) \cdot c_{Cr}$) is added to the local free energy density. This excess energy is shown in Fig. 5b as a function of the order parameter and Cr-concentration. There are a peak and a valley along the line of $\phi=0.4$ and $\phi=0.6$ respectively, and the height and the depth decreasing with decreasing Cr-concentration. The relaxation of static interface was simulated with the same initial condition as in the simulation without segregation energy described in section 2.5.1.

## 3. Result

### 3.1. Equilibrated static interface with only elastic solute-boundary interaction

Figure 6a shows the equilibrated profiles of the order parameter $\phi$ and concentrations of Mo, Nb and Cr near C40/C11$_b$ for the case where Cr-atoms are included as an additive element under the condition described in section *2.5.1*. The results of the cases with and without the elastic strain energy are shown with solid line and dotted line, respectively. There is no significant difference between the results for the cases with and without elastic strain energy. Mo- and Nb-atoms are distributed to both phases but Cr-atoms are distributed mostly in C40-phase. As a result, the volume fraction of C40 phase has increased. Although the distribution of Cr-atoms has dramatically changed from the initial conditions, no segregation is observed at the interface contrary to the experimental result [17]. The Nb-concentration in C11$_b$-phase is slightly lower in the case with elastic strain energy (solid line) than in the case without elastic strain energy (dotted line). But the difference is as small as 0.1 at%. The difference due to the elasticity in the profiles of Cr-concentration is much smaller, and the $c_{Cr}$ is 1.6 at% in C40-phase and 0 at% in C11$_b$-phase in both results. Thus, the effect of elasticity on the solute distribution is negligibly small and it is difficult to attribute the experimentally observed Cr-segregation to the relaxation of lattice misfit. Fig. 6b shows the distribution of elastic strain ($\varepsilon_{11}^{el}$, $\varepsilon_{22}^{el}$ and $\varepsilon_{33}^{el}$). The result without elastic strain energy (dotted lines) is also superimposed for comparison. There is little difference between two results. Elastic strain, which is calculated by Eq. (6), changes continuously in the interface region. The absolute values of elastic strain $\varepsilon_{11}^{el}$, $\varepsilon_{22}^{el}$ and $\varepsilon_{33}^{el}$ are as large as approximately 0.5, 1 and 1 %, respectively, and their variation in the interface region is very steep even after the relaxation. Thus, the relaxation of lattice misfit by the by the solute redistribution is not so significant. This implies that the experimentally observed strong Cr-segregation cannot be attributed to the relaxation of lattice misfit but to another mechanism.

*3.2. Migrating interface with only elastic solute-boundary interaction*

Figure 7a shows the initial distributions of order parameter and solute concentrations for simulating the interface migration whose procedure is in section *2.5.2*. Figs. 7b-d show the distributions of the order parameter $\phi$ and solute concentrations at a moment during the course of interfacial migration simulated with different assumptions of Cr-diffusivity. Volume fraction of C11$_b$-phase has increased as shown in the distribution of the order parameter. It is seen that Nb- and Cr-atoms pile up ahead of the migrating interface. The pile-up of Cr-atoms is much more prominent in the cases of $D_{Cr} = D_{Nb}/2$ and $D_{Cr} = D_{Nb}/8$ than in the case of $D_{Cr} = D_{Nb}$, and becomes more prominent with decreasing $D_{Cr}$. These results indicate that Cr-segregation can take place during the course of solute redistribution due to the interface migration. However, the significant Cr-segregation at the interface gradually diminishes and disappears eventually in equilibrium after solute redistribution completed as in Fig. 6a. It might be suggested that the interface segregation detected experimentally also took place as a result of solute redistribution during the growth of C11$_b$-phase, i.e. the solute piled up at the migrating interface. If this is the case, interfacial segregation would not exist in equilibrium. To address this issue, experiments to measure solute diffusivities, are in progress.

*3.3. Equilibrated static interface with chemical solute-boundary interaction*

Fig. 8a shows the equilibrated distributions of the order parameter $\phi$ and solute concentrations simulated by taking into account the segregation energy as described in section *2.7*.  The results for the two cases with and without elastic strain energy are shown as solid line and dotted line, respectively.  Unlike the result without the segregation energy (Fig. 6a), a slight Cr-segregation is recognized at the interface as indicated by an arrow in both cases with and without elastic strain energy.  However, the Cr-segregation is not so significant compared to that in the experiment where approximately two times higher Cr-concentration is detected at the C40/C11$_b$ lamellar interface than in the C40-and C11$_b$-lamellae.  The difference between the two results with and without the elastic strain energy is negligibly small as well as in the case without the segregation energy (Fig. 6a).  The changes in Nb- and Cr-concentrations due to the elastic strain energy are all less than 0.1 at % in both phases, and the Cr-concentration at the peak of segregation are also nearly equally around 0.9 at %.  Namely, there is no significant effect of elastic strain on the segregation profiles.  Therefore, the segregation energy (i.e. the chemical interaction between solute atoms and the interface) appears more responsible for the segregation rather than the elastic strain energy.  However, the simulated solute segregation is too small to explain the experimentally detected segregation.

## 4. Discussion

The Cr segregation in the simulation (Fig. 8) is much less than that measured in the experiment.  Moreover, the Cr-concentration at the peak of (0.9 at%) is lower than that in C40-phase (1.4 at%) in the simulation whereas the experimentally measured peak concentration is approximately two times as high as that in the C40-matrix.  The composition profile across the C40/C11$_b$ interface was measured by STEM-EDS analysis.  According to the EDS analysis, Cr-concentrations in C40- and $C11_b$-phases are close to each other, and Cr-concentration at the $C40/C11_b$ interface is approximately 2 times higher than that in the C40- and $C11_b$-phase lamellae.  There are two possible reasons for these differences between the simulation and the experiment.  One is the finite resolution of simulation grid, and the other is the contribution of lattice vibration which has recently pointed out by the first principles calculation study [31].  Here their relevance to the inconsistency between the simulation and the experiment is discussed.

*4.1. The resolution of simulation grid*

In this study, the interface thickness $d$ is determined to be 1.3 nm, and the gradient energy coefficients for the order parameter $\kappa_\phi$ and the interfacial-energy density $W_{\text{interface}}$ are estimated from their relationship with the interfacial energy and the using $d$, $\kappa_\phi = (3/4)\gamma_s d$ and $W_{\text{interface}} = \gamma_s/d$.  Considering that there are 6 atomic layers within the $C40/C11_b$ interfacial region, the grid interval must be under $1.3/6 \approx 0.22$ nm to reproduce the segregation at the atomic layers.  In this study, the simulation space with the length of 24 nm is represented by 256 grids, and the grid size is approximately 0.1 nm ($\approx 24/256$ nm).  The grid size seems small enough to reproduce the segregation at the atomic layers.  In order to confirm that the grid size is small enough, additional simulations have been conducted with a variety of grid numbers 128, 256, 512 and 1028 for the fixed space length of 24 nm, and the dependence of the Cr-segregation on the grid number has been examined.  The grid number are chosen from the power-of-two numbers for using Fast Fourier Transform (FFT) which is needed to calculate the elastic strain energy [32, 33].  Figure 9 shows the equilibrated profiles of Cr-concentration around the interfacial region conducted by each condition.  The profile derived from the McLean's equation is also superimposed. The McLean's equation can be expressed as

$$c_{\mathrm{Cr}} = \frac{c_0 \exp(-\Delta g_{\mathrm{gb}}^{\mathrm{seg}}/RT)}{1 + c_0 \exp(-\Delta g_{\mathrm{gb}}^{\mathrm{seg}}/RT)} \tag{15}$$

where $c_0$ denotes the average concentration, and $\Delta g_{\mathrm{gb}}^{\mathrm{seg}}$ denotes the segregation energy [34]. The profile for McLean's equation is obtained by substituting $E_{\mathrm{Cr}}^{\mathrm{seg}}(\phi)$ (Eq.(14)) to $\Delta g_{\mathrm{gb}}^{\mathrm{seg}}$. The segregation profiles for different grid numbers are different from each other. In the system with 128 grid points, the grid size is 0.18 nm (12 nm /64), and it is smaller than the interval between the atomic layers of 0.26 nm (1.3 nm / (6−1)). This grid size may appear small enough to resolve the interfacial region layer by layer. The concentrations in C40- and C11$_b$-phases are same as in the other cases, but the variation within the interfacial region is obviously different, and no segregation is seen at the interface in the 128-grid-point system whereas the segregations are clearly seen in other systems with finer grids. Therefore, this grid size is not small enough to reproduce segregation at the atomic layers. On the other hand, in the system with 512 or 1024 grid points, segregation can be seen in common with the system with 256 grid points though their peak value of Cr segregation are a little larger. Therefore, the grid size of 0.1 nm (grid number of 256) which is used for the main simulation is small enough to reproduce the segregation which can occur on a plane of one atomic layer. The slightly less of Cr-segregation can be due to the lower grid resolution, and a little larger peak occurs by applying much larger grid number (i.e. finer grids). Nonetheless, the difference among the segregation profiles for the 256, 512 and 1024-point-grids are not responsible for the difference between the experiment and the simulation.

*4.2. Effect of vibrational contribution to the segregation energy*

In one of our collaborative researches, the C40/C11$_b$ lamellar interface has been studied by the first principles calculation. Lattice parameters, interfacial energy of C40/C11$_b$ lamellar interface and the segregation energy have been evaluated in the study, and they are used as the input parameters fro the phase-field simulation in the present study.   For conventional alloys, the chemical interaction between the interface and solute atoms can be evaluated by taking into account only electric internal energy. The simulation results in the previous section are also obtained by using the segregation energies evaluated by taking into account only electric internal energy.   Recently, theoretical works revealed that lattice vibration can play a significant role on phase stability in alloys.   The segregation energy was recalculated taking into account the vibrational free energy contributions, and it has been found that the segregation energies of Cr-atoms at the atomic layers within the interface region changes differently with increasing temperature.   This suggests that the temperature affects the segregation energy function of Cr-atoms due to the vibrational contribution.   Therefore the segregation profile may become different when the segregation energy with vibrational contribution is used.

Figure 10 shows the Cr segregation energy (black dots) at 1673 K evaluated by the first principles calculation taking both electronic internal energy and vibrational free energy contributions. The solid line shows the segregation energy function recalculated for 1673 K fitted to the segregation energies at 1673 K. The open circles and the dotted line show the previously used segregation energies and the segregation energy function evaluated with only electronic internal energy (same as in Fig. 8) for comparison.   It is seen that the segregation energies at the layers of $\phi=0.6$ and $\phi=1.0$ are decreased significantly by the lattice vibration, and the decreases are approximately 0.1 eV.   It should be noted that the energy at $\phi=0.6$ is lower than that at $\phi=0$ (C40-matrix).

The solid lines in Fig. 11 show the equilibrated profiles of order parameters and solute concentrations simulated by using the segregation energy with lattice vibrational contribution.   The dotted lines show the results without lattice vibration for comparison (same as in Fig. 7a).   The Cr-segregation is obviously different from that in the case without lattice vibration.   In the system with lattice vibrational effect, strong Cr-segregation is observed, and Cr-concentration is higher at the interface than in the C40-matrix whereas it is lower than in the C40-phase in the case without lattice vibrational effect. Cr-concentration at the interface is nearly two times higher than the average concentration, which is in good agreement with the experiment. This result suggests Cr-segregation at C40-NbSi$_2$/C11$_b$-MoSi$_2$ is thermodynamically stable and is largely dependent on both electronic and lattice vibration contributions of the segregation energy. From the above discussion, the lattice vibrational contribution to the segregation energy is mainly responsible for the difference between Cr-segregation in the experiment and that in the simulation described in the previous section. Nevertheless, the simulation without the lattice vibration is useful to roughly predict the segregation behavior since it is qualitatively agree with the experiment, and the first principles calculation of segregation energy with the lattice vibration contribution is computationally very expensive at this moment.

## 5. Conclusions

A new phase-field model which takes into account the segregation energy has been developed, and the mechanism of Cr-segregation to lamellar interface in C40-NbSi$_2$ / C11$_b$-MoSi$_2$ duplex silicide has been investigated by using the model and parameters provided by first principles calculations. The main findings are as follows.

- Cr-segregation does not occur in the case without segregation energy, and the relaxation of lattice misfit by solute redistribution is negligibly small. Therefore, the Cr-segregation cannot be attributed to the decrease in the elastic strain energy.
- In the case with the segregation energy, a slight Cr-segregation is formed at the static equilibrated interface. However, the Cr-segregation is much less significant than that measured experimentally when the segregation energy evaluated by first principles calculation without lattice vibration is used.
- The profile of Cr-segregation depends on the resolution of grid used in the simulation, and the segregation disappears when a grid coarser than that of this study by a factor of two is used. But the grid used in the main part of present study has been confirmed to be fine enough, and the resolution effect cannot explain the discrepancy between the simulation result and the experimental measurement.

- Significant Cr-segregation comparable to that in the experiment can be formed in during the growth of $C11_b$-lamellae from C40-matrix if the diffusion of Cr is much slower than those of Mo and Nb. But the segregation disappears after equilibration. Also, strong Cr-segregation can be formed in the simulation in which the segregation energy evaluated with lattice vibrational effect taken into account is used.

**Acknowledgments**

This study is supported by the Advanced Low Carbon Technology Research and Development Program of Japan Science and Technology Agency. This study is partly supported by the Center for Computational Materials Science, Institute for Materials Research, Tohoku University and the Cyberscience Center, Tohoku University.

Table 1 Lattice parameters (nm) of each transition metal disilicide evaluated by first principles calculation and lattice sizes (nm) based on Equation (9).

|  | C40 | | $C11_b$ | |
|---|---|---|---|---|
|  | a | c | a | c |
| $MoSi_2$ | 0.4613 | 0.6619 | 0.3217 | 0.8277 |
| $NbSi_2$ | 0.4819 | 0.6627 | 0.3290 | 0.7865 |
| $CrSi_2$ | 0.4404 | 0.6369 | 0.3082 | 0.7534 |

|  | C40 | | | $C11_b$ | | |
|---|---|---|---|---|---|---|
|  | $l_1$ | $l_2$ | $l_3$ | $l_1$ | $l_2$ | $l_3$ |
| $MoSi_2$ | 1.5980 | 1.3839 | 1.3238 | 1.5730 | 1.3649 | 1.3649 |
| $NbSi_2$ | 1.6694 | 1.4457 | 1.3254 | 1.6554 | 1.3958 | 1.3958 |
| $CrSi_2$ | 1.5256 | 1.3212 | 1.2738 | 1.5068 | 1.3076 | 1.3076 |

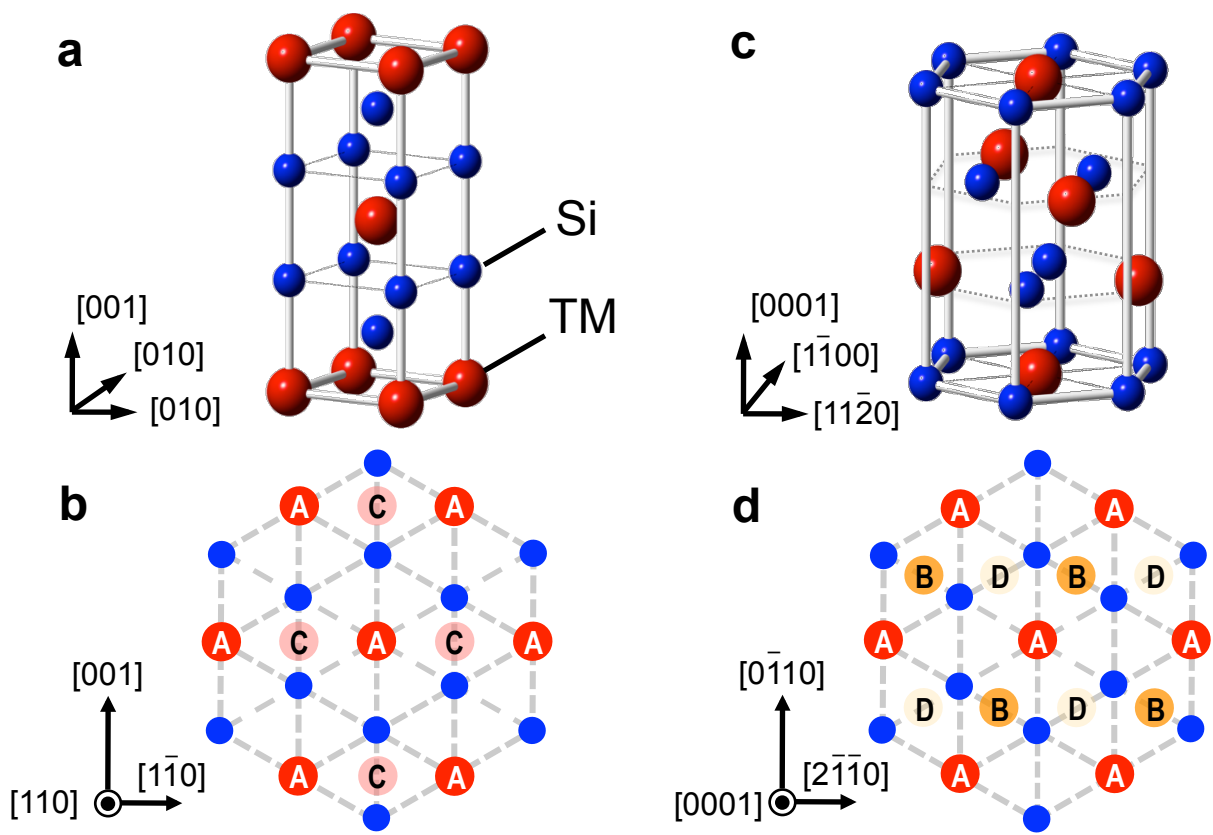

Fig. 1. Unit cells of (a) C40-structure and (b) $C11_b$ structure. Large balls show transition metals (Mo- or Nb-atoms) and small ones show Si-atoms. Atomic arrangements at the lamellar interface of (c) C40 phase and (d) $C11_b$ phase.

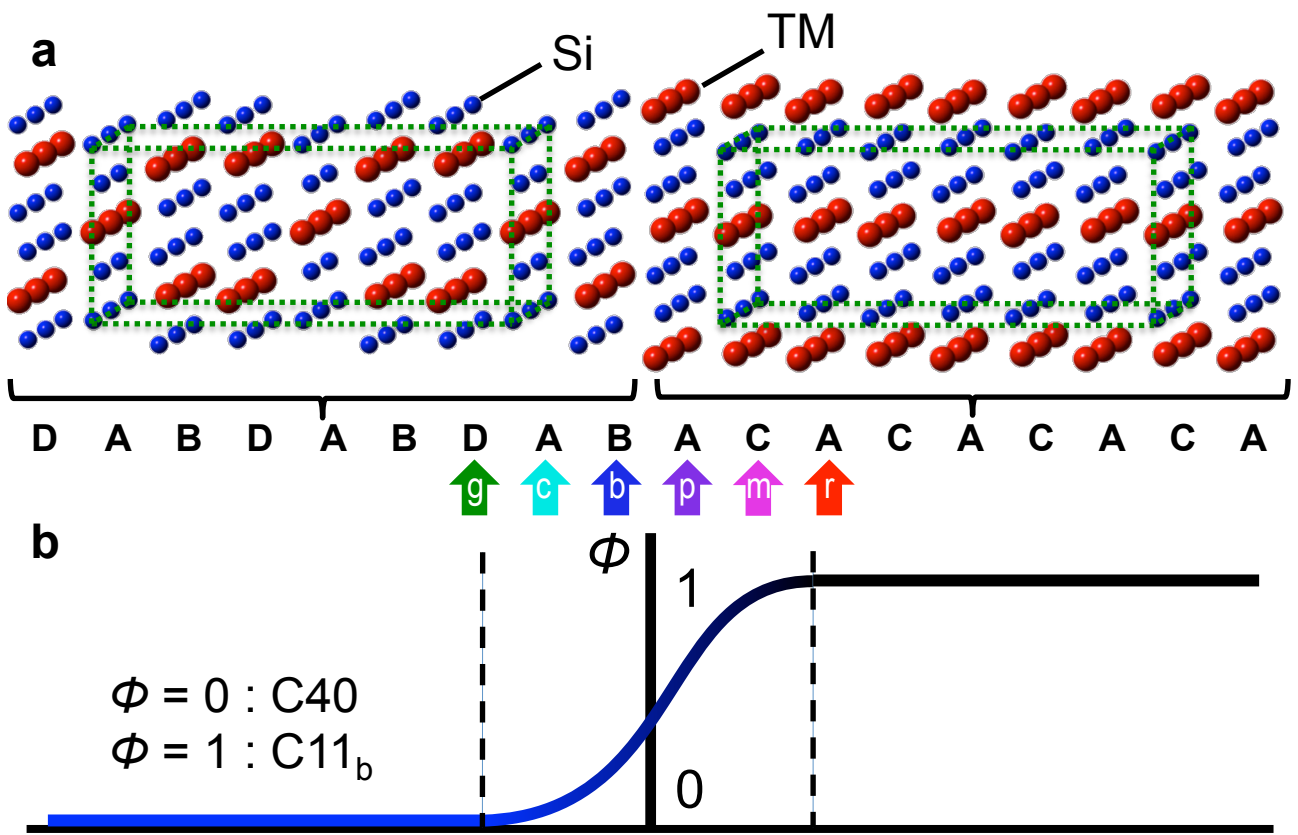

Fig. 2. (a) Atomic layers at the lamellar interface and (b) definition of the order parameter.

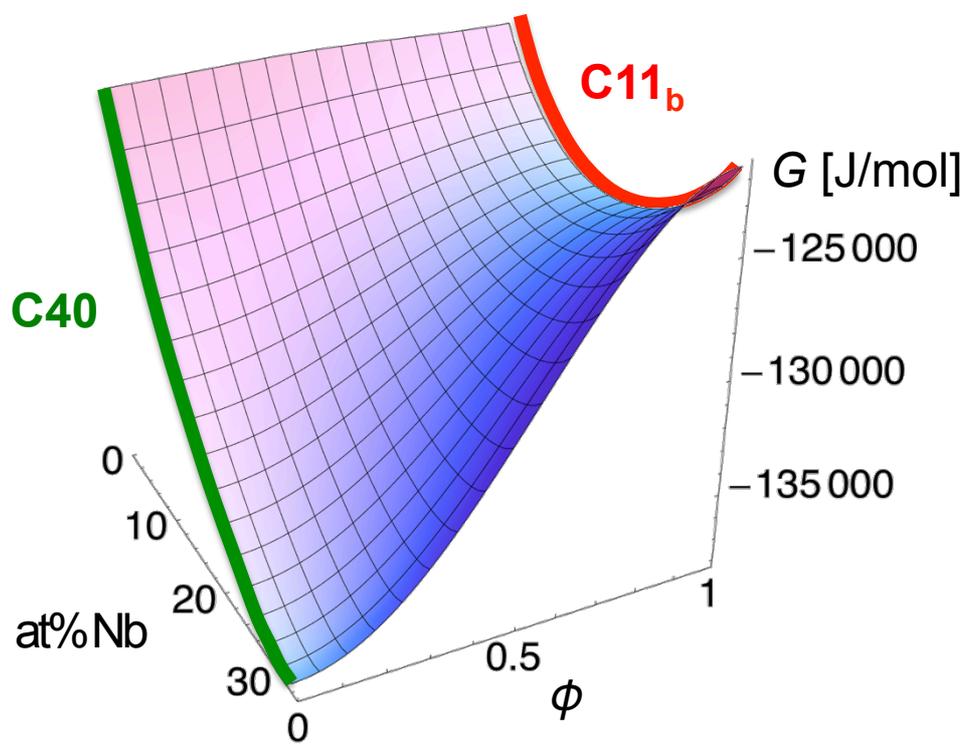

Fig. 3. Landscape of free energy as a function of the order parameter $\phi$ and Nb concentration $c_{Nb}$.

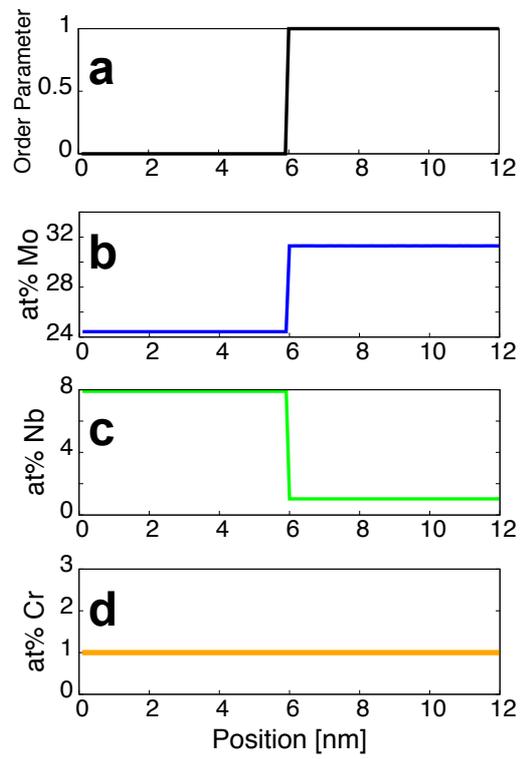

Fig. 4. Initial distributions of (a) the order parameter and solute concentrations of (b) Mo, (c) Nb, (d) Cr across the interface.

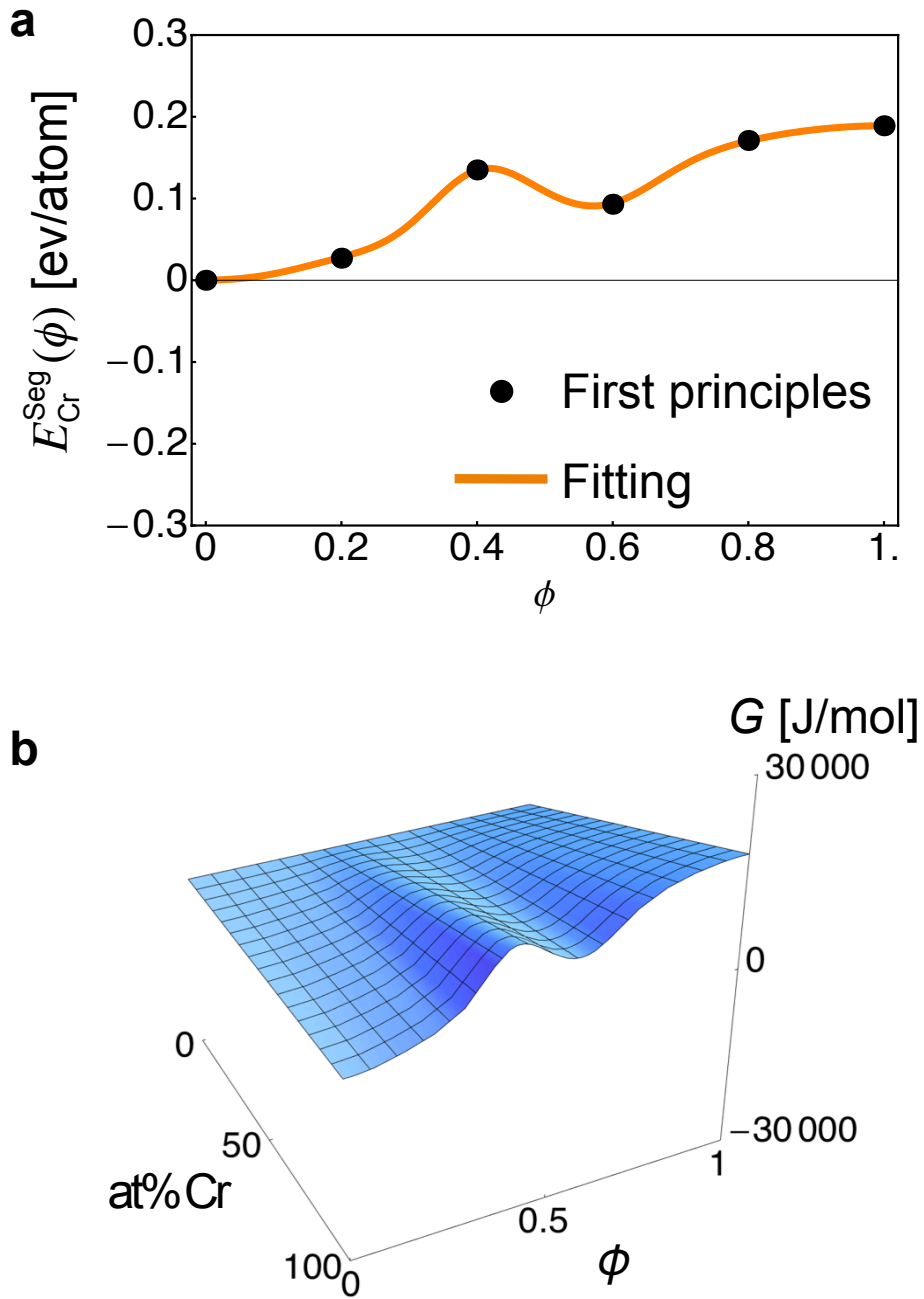

Fig. 5. (a) Segregation energy of each position calculated by first principles calculation (black solid circles) and segregation energy function defined to interpolate the segregation energies. (b) Landscape of segregation energy as a function of the order parameter $\phi$ and additive-element concentration ($c_{Cr}$).

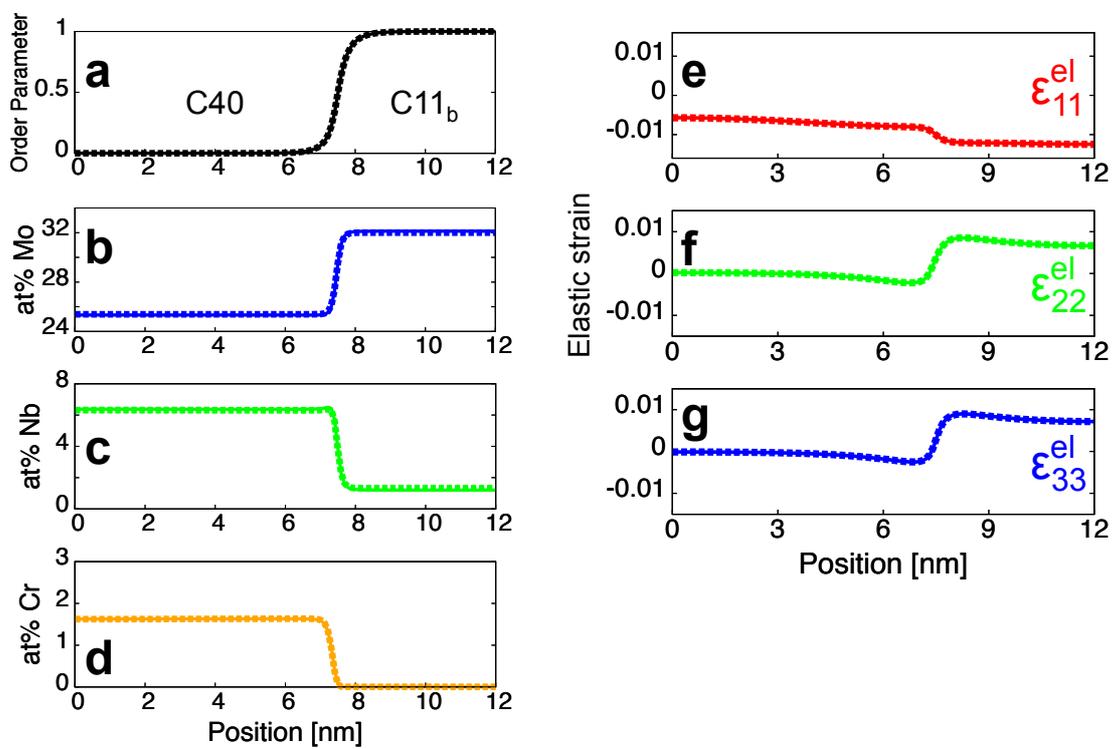

Fig. 6 (a-d) Profiles of the order parameter and solute concentrations of Mo, Nb, Cr across the interface. (e-f) Profiles of elastic strain in each direction. Dotted lines shows the results for the case without elastic strain energy.

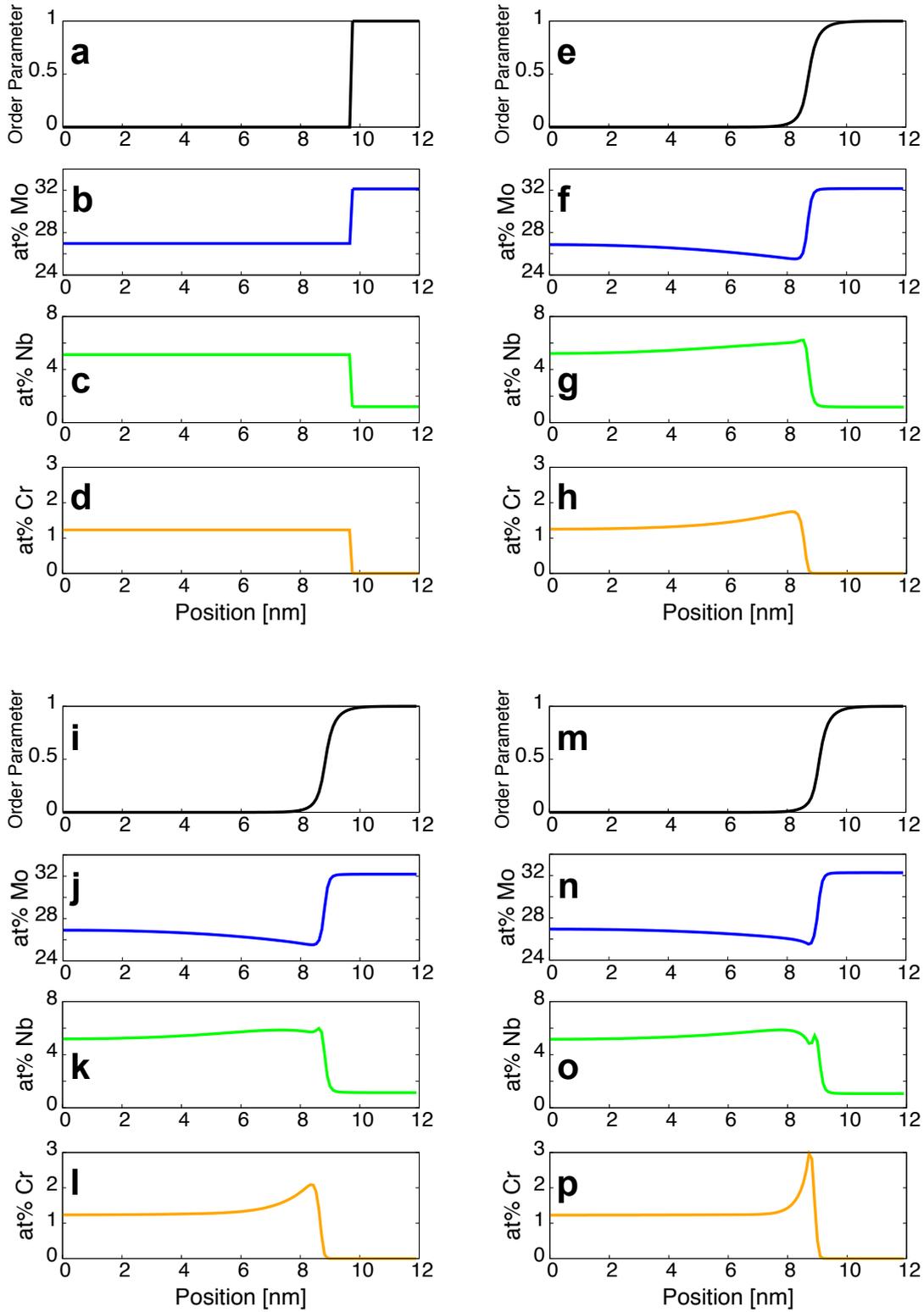

Fig. 7. (a-d) Initial distributions of order parameter and concentrations for the simulation of interface migration. Profiles of order parameter and concentrations around the migrating interface simulated with variously assumed Cr diffusivities: (e-h) $D_{Cr}=D_{Nb}$, (i-l) $D_{Cr}=D_{Nb}/2$ and (m-p) $D_{Cr}=D_{Nb}/8$.

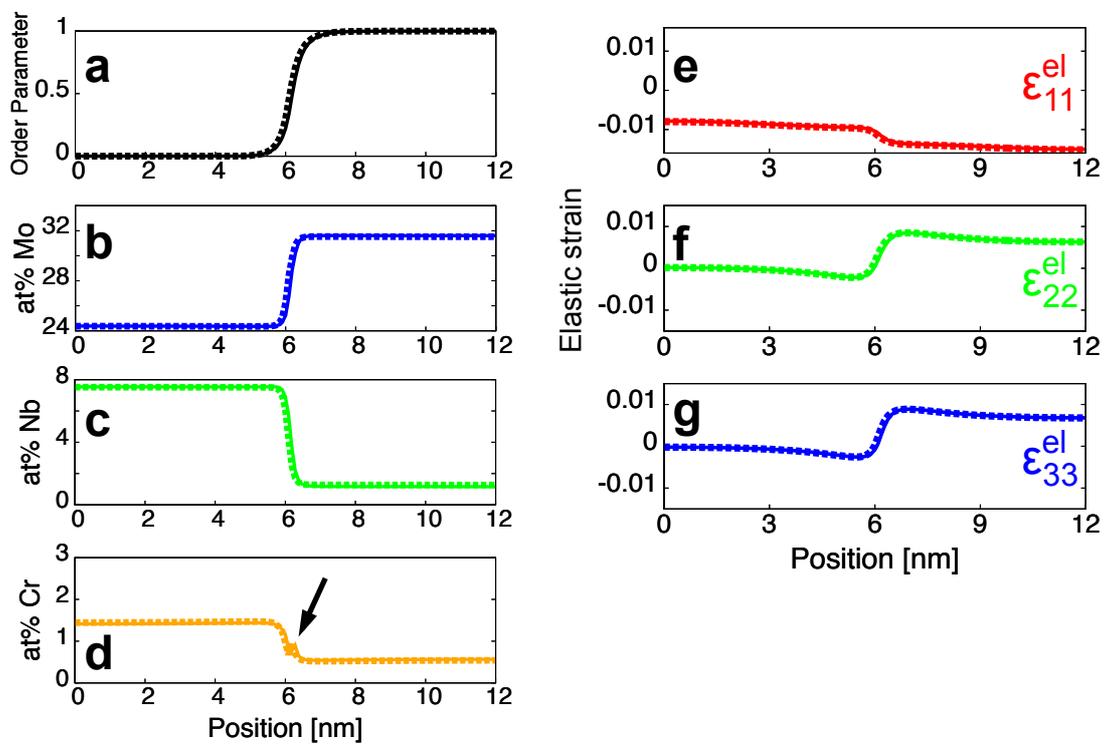

Fig. 8. (a-d) Profiles of the order parameter and solute concentrations simulated by taking into account segregation energy. (e-g) Profiles of elastic strain in each direction. Dotted lines shows the results for the case without elastic strain energy.

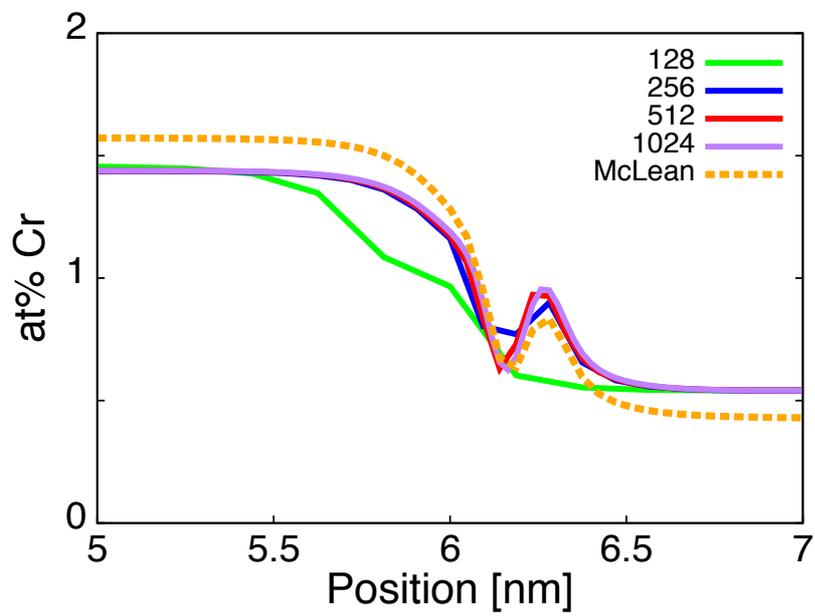

Fig. 9. Profiles of Cr-concentration conducted by various grid number 128, 256, 512 and 1024. Dotted line shows a segregation profile obtained by McLean's equation.

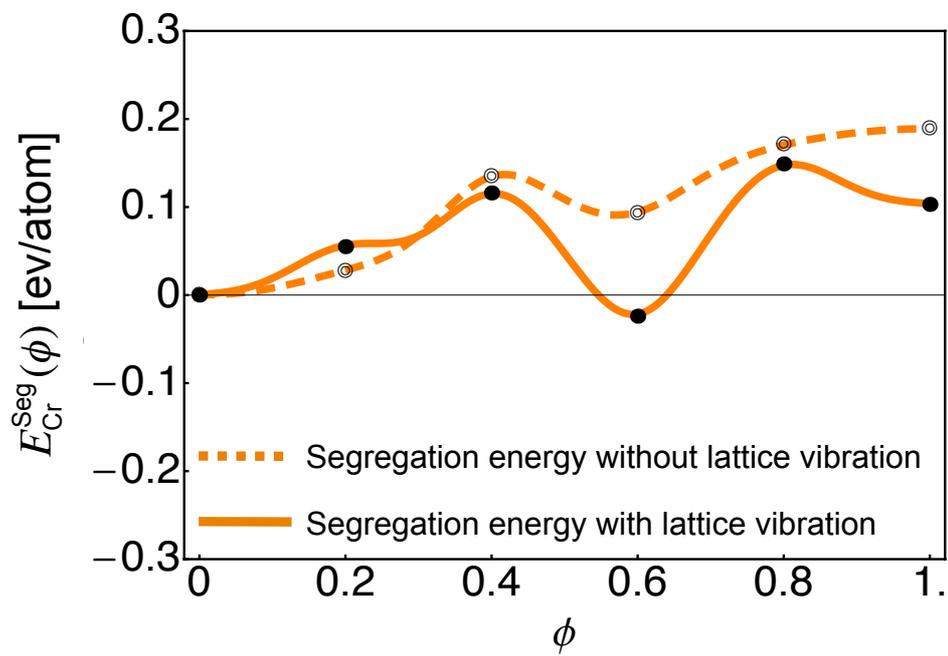

Fig. 10 Solid line shows segregation energy function with lattice vibration. Dotted line shows previous profile for comparison.

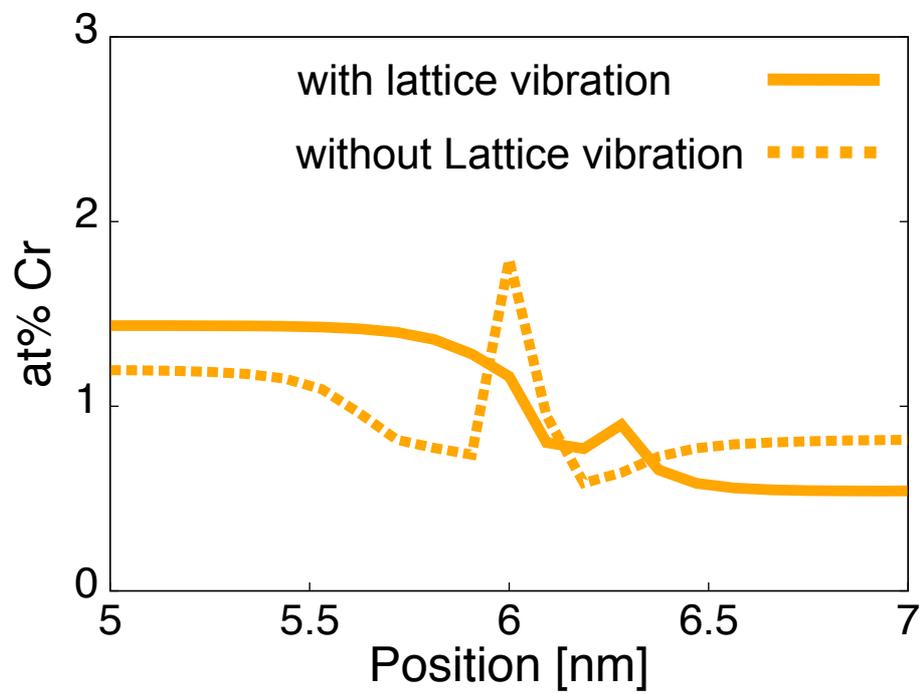

Fig. 11. Profiles of the order parameter and solute concentrations simulated by taking into account segregation energy.